\documentclass{vow2008}

\title{Virtual Observatory studies of Planetary Nebulae.}
\author{R. P. Mignani}
\affil{Mullard Space Science Laboratory, University College London, Holmbury St. Mary, Dorking, Surrey, RH5 6NT, UK}
\author{F. Kerber}
\affil{European Southern Observatory, Karl Schwarzschild Str. 2, Garching, D--85740, Germany}
\author{R. L. Smart}
\affil{INAF, Osservatorio Astronomico di Torino, Via Osservatorio 20, Pino Torinese, I- -10025, Italy}
\author{D. Vande Putte}
\affil{Mullard Space Science Laboratory, University College London, Holmbury St. Mary, Dorking, Surrey, RH5 6NT, UK}
\author{A. Wicenec}
\affil{European Southern Observatory, Karl Schwarzschild Str. 2, Garching, D--85740, Germany}
\author{T. Rauch}
\affil{ Institut fur Astronomie und Astrophysik, Universit\"at T\"ubingen, Sand 1, T\"ubingen, D--72076, Germany}
\author{H.M. Adorf}
\affil{Max Planck Institut fur Astrophysik, Garching, D--85740, Germany}
\author{R. Harrison}
\affil{Rosebury School, Epsom, Surrey, KT18 7NQ, UK }

\usepackage{graphicx}
\begin{document}


\maketitle

\begin{abstract}
Starting from the Strasbourg	ESO Catalogue (SEC) of Planetary Nebulae (PNe), the largest PNe compilation available with 	 $\sim$ 1500 objects, we undertook a comprehensive study of the whole PN population, never carried out so far, only using on-line catalogues and data from public imaging surveys. The study includes the PN dynamics through their measured proper motions (PMs), the study of their galactocentric orbits, the study of their interactions with the interstellar medium (ISM), and the study of their UV-to-IR spectral energy distribution (SED). As a preliminary step required to perform cross-correlations with on-line catalogues, we first went through a systematic reassessment of the PN coordinates (Kerber et al. 2003a). 														
\end{abstract}

\section{Introduction}

About 1500 Planetary Nebulae (PNe) are listed in the Strasbourg ESO Catalogue (SEC) catalogue which represent the largest compilation of PNe produced so far.  We undertook a comprehensive study (morphological classification, proper motions, and colours) of all the PNe in the SEC purely using data available from on-line catalogues. As a preliminary step, we first reassessed the PNe coordinates available in the SEC, affected by random errors as large as several arc minutes, through an interactive comparison with the object coordinates in the {\em GSC-2} reference frame (Kerber et al. 2003a).  This yielded, for all PNe with an identified CS or with a stellar-like morphology, coordinates with an unprecedented accuracy better than 0.3". 
We were thus able to obtain reliable cross-matches between the updated PNe coordinates and those of objects listed in the major all-sky catalogues, to derive multi-band photometry information and proper motion measurements. Moreover, we used the information  on the proper motion to compute, coupled with the distance and radial velocity, the orbits with respect to the galactic center and to investigate the dynamical interactions of the central stars (CSs) of PNe with the interstellar medium (ISM) using imaging data from public H$\alpha$ survey. \\
We stress that this project, the most comprehensive study of PNe so far, has been entirely carried out, and made possible, only using Virtual Observatory data and tools.  In the following sections, we describe the major steps of our projects which have been carried out so far,  and we present early results.  

\section{Proper Motions}

Only a tiny fraction of the about 1500 known Galactic PNe have measured proper motions (PMs). Surprisingly enough, to date, the largest set of PM data for PNe is still the one produced by Cudworth (1974), which includes 62 objects, only 25 of which have PM with significance better than 3$\sigma$ in at least one component. 
We used on-line, all-sky astrometric catalogues available via the Vizier database, i.e. {\em Hipparcos}, {\em Tycho-2}, {\em UCAC2}, {\em USNOB-1.0}, as well as a not yet released version of the {\em GSC-2}, as done in Kerber et al. (2002), to collect PM information for a sample of 1123 stellar-like objects selected from the catalogue of updated  PNe coordinates of Kerber et al. (2003a). We have derived PM information for a total of 234 PNe (274 when including 40 doubtful candidates). We included all PNe for which a confidence level of at least $3 \sigma$ was achieved in at least one PM component. For many objects, PM data are available from more than one catalogue and agreement between different catalogues is usually very reasonable providing independent confirmation of our results. 
We cross-correlated our master catalogue with the CudworthÕs one and we confirmed $\ge 3 \sigma$ PMs for 10 objects and we found $\ge 3 \sigma$ PMs for another 5 objects with $\le 3 \sigma$ PMs in the CudworthÕs catalogue. With 234 (274 with doubtful candidates) entries, we thus produced the largest available compilation of PMs for PNe and their central stars (CSs), which enlarges by almost an order of magnitude the number of PNe and CSs with reliably measured PM (i.e.better than $3 \sigma$ in at least one component). 
This compilation (Kerber et al. 2008) provides us with the unique opportunity of carrying out a systematic study, for the first time on a large sample, of the galactocentric orbits of PNe, following the test case of Sh 2-68 (Kerber et al. 2004), as well as of carrying out a systematic study of the dynamical interactions between the PN and the Interstellar Medium (ISM).

\section{Interaction with the ISM}

We used images of the fields of all PNe with measured PMs (see previous section) obtained  from public H$\alpha$ surveys to search for trails or extended structures which bring evidence of dynamical interactions between the PN and the ISM. For the southern hemisphere we used data from the  {\em Southern H-Alpha Sky Survey Atlas}  
({\em SHASSA}) which covers the sky at  $\le 15^{\circ}$  with $13^{\circ} \times 13^{\circ}$ tiles (0.48Ó/pixel spatial resolution), while for northern hemisphere we used the {\em Virginia Tech Spectral-Line Survey} ({\em VTSS}) which covers the sky at $\ge 15^{\circ}$  with $10^{\circ} \times 10^{\circ}$ tiles (1.6Õ/pixel spatial resolution).   By cross-correlating our  PNe PM compilationd (Kerber et al. 2008) with the {\em SHASSA} and {\em VTSS} fields we found matches in either surveys for 224 PNe. After filtering out  PNe with doubtful PMs we ended up with a clean working sample of 198 objects. We retrieved  the H$\alpha$ images from the {\em SHASSA} and {\em VTSS} web sites and we visually inspected each field by overlaying the PN coordinates and PM vectors. Each  dataset was screened to search  for extend cometary-like trails and for other  interesting  features   which  might  be   associated  with  dynamical interactions between the PNe and the ISM like, e.g.  bow-shocks.   As a  safe measure, the data  screening  was repeated  at  least  two  times and  carried  out
independently by three members of our team to have a better handle on bias effects. Although the clearest case of an H$\alpha$ trail is that of the PN Sh 2-68 (Kerber et al. 2003b), we found a few more interesting cases (Kerber et al. 2009) which we are now investigating through  dedicated follow-up observations.

\section{Spectral energy distribution}

Starting from the sample of 1123 stellar-like PNe (Kerber et al. 2003a), we have cross-matched their positions with on-line, all-sky object catalogues to build up their ultaviolet-to-infrared spectral energy distribution (SED). In the optical, we have used as a reference  the {\em GSC-2}\footnote{ttp://www-gsss.stsci.edu/Catalogs/GSC/GSC2/GSC2.htm} catalogue which provides photometry information in four photographic passbands (overlaying the B,V,R,I Johnson passbands). In the near-infrared (NIR), we used both the DENIS\footnote{http://www-denis.iap.fr/}  survey (I, J, and K bands), and the 2MASS\footnote{http://www.ipac.caltech.edu/2mass/} survey (J, H, and Ks). In the ultraviolet (UV), photometry data in broad near-UV (NUV) and far-UV (FUV) bands have been obtained from the GALEX\footnote{http://galex.stsci.edu/GR1/}  survey and, in a few cases only, from the recently releases {\em XMM/Optical Monitor} catalogue  (Still et al., these proceedings) which provides deeper photometry in the Johnson B and V bands with respect to the {\em GSC-2}, as well as photometry in the U and in two NUV bands.  We performed object matching using available VO tools for catalogue cross-matching. We then derived the observed SEDs from the spectral fluxes computed from the available filter curves. For the {\em GSC-2}, we empirically computed the conversion to the Johnson system  by computing the color transformation by cross-matching photometric standards in the Stetson's catalogue. We  corrected the observed SEDs  for customized values of the  interstellar extinction towards the PN, as derived from values compiled from the literature. We then compared the  extinction-corrected SEDs  with theoretical ones simulated from available stellar spectral libraries, using ad hoc tools (Adorf et al. 2005) developed within the German Astronomical Virtual Observatory (GAVO)\footnote{http://www.g-vo.org/www/} project and we assessed the reliability of our results on a number of test cases (Kerber et al., in preparation).  



\end{document}